\documentclass[conference]{IEEEtran}
\IEEEoverridecommandlockouts
\usepackage{cite}
\usepackage{amsmath,amssymb,amsfonts}
\usepackage{algorithmic}
\usepackage{graphicx}
\usepackage{textcomp}
\usepackage{xcolor}

\def\BibTeX{{\rm B\kern-.05em{\sc i\kern-.025em b}\kern-.08em
    T\kern-.1667em\lower.7ex\hbox{E}\kern-.125emX}}
\begin{document}

\title{A Gray-Box Stability Analysis Mechanism for Power Electronic Converters\\
}

\author{\IEEEauthorblockN{ Rui Kong, Subham Sahoo, Yubo Song, and Frede Blaabjerg}
\IEEEauthorblockA{\textit{Department of Energy} \\
\textit{Aalborg University}\\
Aalborg, Denmark \\
ruko@energy.aau.dk, sssa@energy.aau.dk, yuboso@energy.aau.dk, fbl@energy.aau.dk}

}

\maketitle

\begin{abstract}
This paper proposes a gray-box stability analysis mechanism based on data-driven dynamic mode decomposition (DMD) for commercial grid-tied power electronics converters with limited information on its control parameters and topology. By fusing the underlying physical constraints of the state equations into data snapshots, the system dynamic state matrix and input matrix are simultaneously approximated to identify the dominant system dynamic modes and eigenvalues using the DMD with control (DMDc) algorithm. While retaining the advantages of eliminating the need for intrinsic controller information, the proposed gray-box method establishes higher accuracy and interpretable outcomes over the conventional DMD method. Finally under experimental conditions of a low-frequency oscillation scenario in electrified railways featuring a single-phase converter, the proposed gray-box DMDc is verified to identify the dominant eigenvalues more accurately.
\end{abstract}


\section{Introduction}
With extensive integration of grid-tied power electronics converters, various instability issues occur \cite{r1}. To assess such unstable scenarios, many model-based analysis approaches are proposed, such as the small-signal impedance method \cite{r2} and the eigenvalue analysis method \cite{r3}, which complies with detailed controller information as a mandatory pre-requisite, resulting in great difficulties due to the low transparency of commercial power electronic converters. On the other hand, data-driven approaches are feasible for black-box conditions that consequently lack interpretability in general \cite{r4}. 

As a burgeoning data-driven technique, dynamic mode decomposition (DMD) has been widely applied in fluid dynamics and other fields \cite{r5,r6}, attempting to find the best fit dynamic matrix from the measured data snapshots. It has the capability to identify dynamic modes and eigenvalues for system stability assessment with underlying properties of each mode like damping and frequency. DMD is extensively used for oscillation analysis in power systems \cite{r7,r8}, providing better performance over other data-driven algorithms such as Prony \cite{r9} and eigensystem realization \cite{r10} in terms of accuracy and identification of the spatio-temporal characteristics. However, the limited number of measurable states of converters cannot meet the high dimension precondition of data matrices to ensure containment of all the significant modes \cite{r11}. Moreover, there is no clear standard behind the selection of measured states, resulting in the inclusion of all measurable signals in a pair of data matrices for approximating the dynamic matrix without any regard for the semantic co-relationship between physics and data. With power electronic converters being actuated systems not only having inherent dynamics but also external inputs, the accuracy of DMD-based stability analysis will be significantly affected when only unforced dynamics are considered \cite{r12}. 

To address the above-mentioned problems, a DMD-based gray-box stability analysis mechanism is proposed in this paper considering the physical constraint extracted from the converter state equations. The improved dynamic mode decomposition with control (DMDc) algorithm is employed considering the impact of external inputs. Furthermore, the data-stacking technique is used to generalize the converter system dynamic with few measurement channels. Based on the experimental platform of a single-phase converter under the low-frequency oscillation scenario of electrified railways, it is verified that the proposed method can assess stability by identifying oscillation modes with more accuracy and enhanced interpretability.

\section{Dynamic Mode Decomposition for Stability Analysis}

\begin{figure}[htbp]
\centering
\includegraphics[width=0.9\columnwidth]{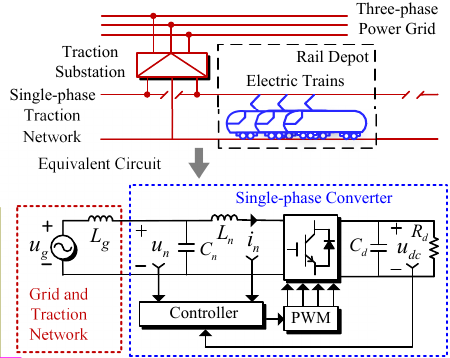}
\caption{System diagram of a single-phase converter in electrified railways.}
\label{Fig1}
\end{figure}

Considering the single-phase converter of electric trains under low power conditions as shown in Fig. \ref{Fig1}, it may lead to low-frequency oscillations in electrified railways when multiple trains are energized in the same rail depot \cite{r13}. The state equations of the converter main circuit can be given by:


\begin{equation}
\label{Equ1}
\left[ {\begin{array}{*{20}{c}}
{\dot {i_n}}\\
{\dot{u_{dc}}}
\end{array}} \right] = \underbrace {\left[ {\begin{array}{*{20}{c}}
{ - \dfrac{{{R_n}}}{{{L_n}}}}&{ - \dfrac{{{d_n}}}{{{L_n}}}}\\
{\dfrac{{{d_n}}}{{{C_d}}}}&{ - \dfrac{1}{{{R_d}{C_d}}}}
\end{array}} \right]}_A\left[ \begin{array}{l}
{{i}_n}\\
{u_{dc}}
\end{array} \right] + \underbrace { \begin{bmatrix}
\dfrac{1}{{{L_n}}}\\
{0}
\end{bmatrix}}_B{u_n}
\end{equation}

where, ${R_n}$ and ${L_n}$ are AC side resistance and inductance, ${C_d}$ and ${R_d}$ are DC side capacitance and resistance, ${u_n}$ and ${i_n}$ are AC-side voltage and current, ${u_{dc}}$ is DC-side voltage, and ${d_n}$ is the switching state from the PWM stage.

In the traditional impedance method, the small-signal impedance model of the converter can be established based on the system state equation, shown in (1). Then, an impedance ratio transfer function is formulated for stability assessment by combining converter impedance with grid impedance \cite{r2, r13}, where a thorough derivation of ${d_n}$ involving all control loops is essential. However, detailed controller information for commercial converters is typically unknown.

\begin{figure}[htbp]
\centering
\includegraphics[width=0.9\columnwidth]{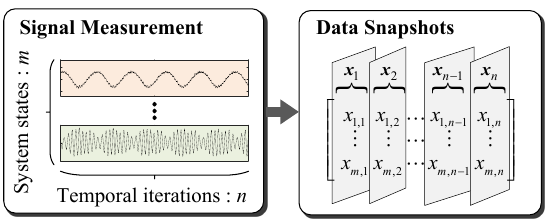}
\caption{Illustration of data collection and data snapshot generation.}
\label{Fig2}
\end{figure}

As a powerful data-driven method for dynamical systems analysis, dynamic mode decomposition (DMD) can find the best fit dynamic matrix using the spatio-temporal snapshots of measurement data to extract dynamic modes and eigenvalues for system stability analysis. We assume that ${x_{i,j}}$ denotes an element of the measurable discrete system states as:
\begin{equation}
\label{Equ2}
{x_{i,j{\rm{ }}(\begin{array}{*{20}{c}}
{i = 1,2, \ldots ,m,}&{j = 1,2, \ldots ,n}
\end{array})}} = {x_i}(j\Delta t)
\end{equation}
where, $m$ is the number of spatial system states, and $n$ refers to the temporal iteration with regular sampling time interval $\Delta t$.
Therefore, as shown in Fig. \ref{Fig2}, the data snapshots in time can be generated from the data matrix of the discrete dynamical system as:
\begin{equation}
\label{Equ3}
{\bf{X}}_1^n = \left[ {\overbrace {\begin{array}{*{20}{c}}
{{x_{1,1}}}\\
 \vdots \\
{{x_{m,1}}}
\end{array}}^{{{\bf{x}}_{1}}}\overbrace {\begin{array}{*{20}{c}}
{{x_{1,2}}}\\
 \vdots \\
{{x_{m,2}}}
\end{array}}^{{{\bf{x}}_{2}}}\begin{array}{*{20}{c}}
 \cdots \\
 \cdots \\
 \cdots 
\end{array}\overbrace {\begin{array}{*{20}{c}}
{{x_{1,n - 1}}}\\
 \vdots \\
{{x_{m,n - 1}}}
\end{array}}^{{{\bf{x}}_{{n - 1}}}}\overbrace {\begin{array}{*{20}{c}}
{{x_{1,n}}}\\
 \vdots \\
{{x_{m,n}}}
\end{array}}^{{{\bf{x}}_{n}}}} \right]
\end{equation}
where, the data matrix ${\bf{X}}_1^{n} \in {\mathbb{R}^{m \times n}}$ consists of $n$ snapshots $({{\bf{x}}_{1}}{\bf{,}}{{\bf{x}}_{2}},...,{{\bf{x}}_{n}})$, and the subscript and superscript represent the first and last measurement snapshots in the matrix, respectively. 

For a discrete-time linear system, its dynamical evolution is governed by the mapping between adjacent snapshots shown as:
\begin{equation}
\label{Equ4}
\begin{array}{*{20}{c}}
\begin{array}{*{20}{c}}
{{{\bf{x}}_{{j + 1}}} = {\bf{A}}{{\bf{x}}_{j}},}\ {\rm{ }}{( j = 1,2...,n - 1 )}
\end{array}
\end{array}
\end{equation}
where, ${\bf{A}} \in {\mathbb{R}^{m \times m}}$ is the dynamic matrix capturing the inherent dynamics in the data matrix. It is worth mentioning that the approximated operator in (4) can also be used on data generated by non-linear dynamical systems \cite{r14}. 

To this end, the DMD algorithm centers around finding a best-fit solution of the matrix ${\bf{A}}$ for each pair of measurements. Firstly, the measured state snapshots are gathered to generate a pair of data matrices, ${\bf{X}}_1^{n - 1} \in {\mathbb{R}^{m \times (n - 1)}}$ and ${\bf{X}}_2^n \in {\mathbb{R}^{m \times (n - 1)}}$, where ${\bf{X}}_1^{n - 1}$ has a time shift ahead of ${\bf{X}}_2^n$, which can be shown as:

\begin{equation}
\label{Equ5}
\left\{ \begin{array}{l}
{\bf{X}}_1^{n - 1} = \left[ {\begin{array}{*{20}{c}}
{{{\bf{x}}_1}}&{{{\bf{x}}_2}}& \ldots &{{{\bf{x}}_{n - 1}}}
\end{array}} \right]\\
\begin{array}{*{20}{c}}
{}&{}&{ = \left[ {\begin{array}{*{20}{c}}
{{{\bf{x}}_1}}&{{\bf{A}}{{\bf{x}}_1}}& \ldots &{{{\bf{A}}^{n - 2}}{{\bf{x}}_1}}
\end{array}} \right]}
\end{array}\\
{\bf{X}}_2^n = \left[ {\begin{array}{*{20}{c}}
{{{\bf{x}}_2}}&{{{\bf{x}}_3}}& \ldots &{{{\bf{x}}_n}}
\end{array}} \right]\\
\begin{array}{*{20}{c}}
{}&{ = \left[ {\begin{array}{*{20}{c}}
{{\bf{A}}{{\bf{x}}_1}}&{{{\bf{A}}^2}{{\bf{x}}_1}}& \ldots &{{{\bf{A}}^{n - 1}}{{\bf{x}}_1}}
\end{array}} \right]}
\end{array}
\end{array} \right.
\end{equation}

It can be seen from (5) that:
\begin{equation}
\label{Equ6}
{\bf{AX}}_1^{n - 1} = {\bf{X}}_2^n
\end{equation}

As a result, the dynamic matrix ${\bf{A}}$ is obtained as:
\begin{equation}
\label{Equ7}
{\rm{ }}{\bf{A}} = {\bf{X}}_2^n{({\bf{X}}_1^{n - 1})^\dag }
\end{equation}
where, $\dag $ denotes the Moore-Penrose pseudo-inverse. 

\begin{figure*}[htbp]
\centering
\includegraphics[width=1.95 \columnwidth]{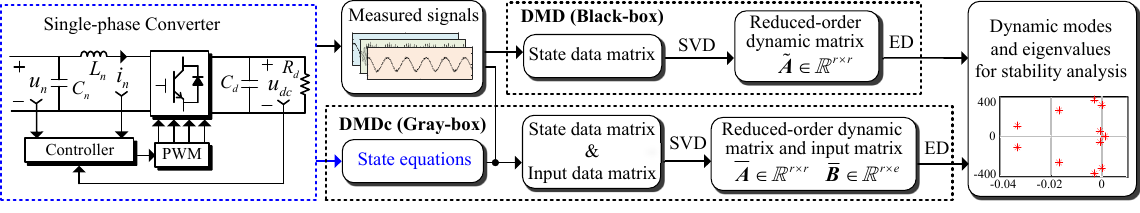}
\caption{Schematic of black-box dynamic mode decomposition (DMD) and proposed gray-box dynamic mode decomposition with control (DMDc) method (SVD: singular value decomposition, ED: eigen-decomposition).}
\label{Fig3}
\end{figure*}

The pseudo-inverse computation of dynamic matrix ${\bf{A}}$ is intractable due to the large dimension of the data matrix ${{\bf{X}}_1^{n - 1}}$. Therefore, in the DMD algorithm, a computationally efficient method through the singular value decomposition (SVD) is implemented to calculate ${({\bf{X}}_1^{n - 1})^\dag }$ with the truncation rank $r$ ${\rm{ }}(r \ll m)$ as: 
\begin{equation}
\label{Equ8}
{\bf{X}}_1^{n - 1} \approx {\bf{U\Sigma }}{{\bf{V}}^T}
\end{equation}
where, ${\bf{U}} \in {\mathbb{R}^{m \times r}}$ and ${\bf{V}} \in {\mathbb{R}^{(n - 1) \times r}}$ are unitary matrices, ${\bf{\Sigma }} \in {\mathbb{R}^{r \times r}}$ is the diagonal matrix of singular values. Then, by projecting the initial full dynamic matrix ${\bf{A}}$ onto an ${\bf{U}}$ basis subspace of dimension $r$, a reduced-order matrix ${\bf{\tilde A}} \in {\mathbb{R}^{r \times r}}$ can be obtained by combining with (7) and (8) as:
\begin{equation}
\label{Equ9}
{\bf{\tilde A}} = {{\bf{U}}^T}{\bf{AU}} = {{\bf{U}}^T}{\bf{X}}_2^n{\bf{V}}{{\bf{\Sigma }}^{ - 1}}
\end{equation}

The next step is to compute the eigen-decomposition (ED) of ${\bf{\tilde A}}$ as:
\begin{equation}
\label{Equ10}
{\bf{\tilde A}} = {\bf{W\Lambda }}{{\bf{W}}^{ - 1}}
\end{equation}
where, ${\bf{W}}$ is the eigenvectors of ${\bf{\tilde A}}$, and ${\bf{\Lambda }} = diag\left[ {\begin{array}{*{20}{c}} {{\lambda _1}}&{{\lambda _2}}& \cdots &{{\lambda _r}} \end{array}} \right]$ consisting of eigenvalues, which can be employed for stability analysis revealing fundamental properties of system with different damping and frequencies. 

The eigenvalues of ${\bf{A}}$ and ${\bf{\tilde A}}$ are equivalent, but the higher computation efficiency of eigen-decomposition of ${\bf{\tilde A}}$ is the crucial advantage of the DMD algorithm, where the number of eigenvalues to be extracted is reduced from $m$ to $r$. In addition, the eigenvectors of ${\bf{A}}$ is defined as DMD exact modes ${\bf{\Phi }}\in {\mathbb{R}^{m \times r}}$, indicating that there is ${\bf{A}} = {\bf{\Phi \Lambda }}{{\bf{\Phi }}^{ - 1}}$. Therefore, the relationship between the ${\bf{\Phi }}$ and ${\bf{W}}$ can be shown as:
\begin{equation}
\label{Equ11}
{\bf{\Phi }} = {\bf{X}}_2^n{\bf{V}}{{\bf{\Sigma }}^{ - 1}}{\bf{W}}
\end{equation}

Furthermore, the measured signal can be reconstructed by combining DMD exact modes with (4) to verify the implementation of the algorithm, shown as:
\begin{equation}
\label{Equ12}
{{\bf{x}}_{j + 1}} = {\bf{\Phi \Lambda }}{{\bf{\Phi }}^{ - 1}}{{\bf{x}}_j} = {\bf{\Phi }}{{\bf{\Lambda }}^j}\underbrace {{{\bf{\Phi }}^{ - 1}}{{\bf{x}}_1}}_{\bf{b}} = \sum\limits_{k = 1}^r {{{\bf{\Phi }}_k}} {({\lambda _k})^j}{b_k}
\end{equation}
where, ${\bf{b}}$ is defined as the mode amplitude, depending on the initial measurement snapshot. ${{\bf{\Phi }}_k}$ is the $k^\text{th}$ column of matrix ${\bf{\Phi }}$, ${\lambda _k}$ is the $k^\text{th}$ eigenvalue, ${b_k}$ is the $k^\text{th}$ mode amplitude for $k=1,2,...,r$.

Finally, the integral contribution $IC$ reflecting the importance of each eigenvalue can be calculated as shown in (13) to determine the dominant mode \cite{r15}:
\begin{equation}
\label{Equ13}
I{C_i} = \Delta t\left\| {{{\bf{\Phi }}_i}} \right\|_F^2\sum\limits_{j = 1}^n {\left| {{{({\lambda _i})}^j}{b_i}} \right|} 
\end{equation}
where, ${\left\|  \cdot  \right\|_F}$ is the Frobenius norm.

DMD has the potential to be a useful diagnostic tool for system stability, but it simply collects all measurable signals to generate state data matrices for approximating dynamic matrix ${\bf{A}}$ without considering the inherent physical structure of the converter system. 

\section{Improved Gray-Box Stability Analysis Mechanism}
The physical constraint extracted from the state equations in (1) of the converter main circuit shows that ${i_n}$ and ${u_{dc}}$ are system states, while ${u_n}$ should be regarded as external input. At this point, DMD considers only the unforced dynamic matrix ${\bf{A}}$ without the input matrix ${\bf{B}}$, which may decrease the evaluation accuracy of the system stability. 

\subsection{Stability Analysis with Gray-Box DMDc}
A gray-box method combining the physical constraint and dynamic mode decomposition with control (DMDc) algorithm is proposed to tackle the aforementioned problems. DMDc algorithm is employed to find the best fit dynamic matrix ${\bf{A}}$ and input matrix ${\bf{B}}$ of the system based on a trio of measurement data matrices, given as:
\begin{equation}
\label{Equ14}
{\bf{X}}_2^n = [\begin{array}{*{20}{c}}
{\bf{A}}&{\bf{B}}
\end{array}]{[\begin{array}{*{20}{c}}
{{\bf{X}}_1^{n - 1}}&{{\bf{U}}_1^{n - 1}}
\end{array}]^{\rm T}} \buildrel \Delta \over = {\bf{G\Omega }}_1^{n - 1}
\end{equation}
where, ${\rm T}$ denotes the transpose, the definitions of state data matrices ${\bf{X}}_1^{n - 1},{\bf{X}}_2^n \in {\mathbb{R}^{m \times (n - 1)}}$ are the same as (5), ${\bf{U}}_1^{n - 1} \in {\mathbb{R}^{e \times (n - 1)}}$ is the data matrix of external inputs, and $e$ is the total number of inputs. Further, matrix ${\bf{\Omega }}_1^{n - 1} \in {\mathbb{R}^{(m + e) \times (n - 1)}}$ is formed by vertically merging ${\bf{X}}_1^{n - 1}$ and ${\bf{\Omega}}_1^{n - 1}$. According to (14), the composite matrix ${\bf{G}}$ is obtained as:
\begin{equation}
\label{Equ15}
{\bf{G}} = {\bf{X}}_2^n{({\bf{\Omega }}_1^{n - 1})^\dag }
\end{equation}

In analogy with DMD algorithm, SVD is exploited with the truncation rank $p$ ${\rm{ }}(p \ll m)$ to compute ${({\bf{\Omega }}_1^{n - 1})^\dag }$  as:
\begin{equation}
\label{Equ16}
{\bf{\Omega }}_1^{n - 1} \approx {{\bf{U}}_p}{{\bf{\Sigma }}_p}{\bf{V}}_p^{\rm T}
\end{equation}
where, the definitions of all notations are the same as (8), and subscript $p$ is used to distinguish them. However, ${{\bf{U}}_p}$ cannot be used to define the subspace on which the state evolves \cite{r12}, because ${{\bf{U}}_p} \in {\mathbb{R}^{(m + e) \times p}}$ involves two components ${{\bf{U}}_{p1}} \in {\mathbb{R}^{m \times p}}$ and ${{\bf{U}}_{p2}} \in {\mathbb{R}^{e \times p}}$, which correspond to ${\bf{X}}_1^{n - 1}$ and ${\bf{U}}_1^{n - 1}$ respectively. As a result, the second SVD of ${\bf{X}}_2^n$ with the truncation rank $r$ $(r \le p)$ is required as:
\begin{equation}
\label{Equ17}
{\bf{X }}_2^n  \approx {{\bf{U}}_r}{{\bf{\Sigma }}_r}{\bf{V}}_r^{\rm T}
\end{equation}

Then, by projecting the initial full dynamic matrix ${\bf{A}} \in {\mathbb{R}^{m \times m}}$ and input matrix ${\bf{B}} \in {\mathbb{R}^{m \times e}}$ into the ${\bf{U}}_r \in {\mathbb{R}^{m \times r}}$ basis subspace, reduced-order matrices ${\bf{\bar A}} \in {\mathbb{R}^{r \times r}}$ and ${\bf{\bar B}} \in {\mathbb{R}^{r \times e}}$ can be given by combining with (15) and (16) as:
\begin{equation}
\label{Equ18}
\left\{ \begin{array}{l}
{\bf{\bar A}} = {\bf{U}}_r^T{\bf{A}}{{\bf{U}}_r} = {\bf{U}}_r^T{\bf{X}}_2^n{{\bf{V}}_p}{\bf{\Sigma }}_p^{ - 1}{\bf{U}}_{p1}^T{{\bf{U}}_r}\\
{\bf{\bar B}} = {\bf{U}}_r^T{\bf{B}} = {\bf{U}}_r^T{\bf{X}}_2^n{{\bf{V}}_p}{\bf{\Sigma }}_p^{ - 1}{\bf{U}}_{p2}^T
\end{array} \right.
\end{equation}

Subsequent steps of DMDc are similar to DMD. Eigenvectors and eigenvalues can be obtained by computing the ED of ${\bf{\bar A}}$ for stability assessment. DMD exact modes can be found, while integral contribution can be calculated using (13) and measured signals are reconstructed for algorithm verification.

The comparison between DMD and DMDc algorithm for stability analysis is shown in Fig. \ref{Fig3}, where the state equations in (1) are utilized in the DMDc algorithm to provide a physical basis for the selection of not only state measurements but also input measurements for (14). Therefore, compared with the conventional DMD algorithm, both the underlying system dynamics and the effects of inputs are involved in the DMDc algorithm, which formulates a more accurate model with interpretability features. 

\subsection{Data Stacking}
\begin{figure}[htbp]
\centering
\includegraphics[width=0.95\columnwidth]{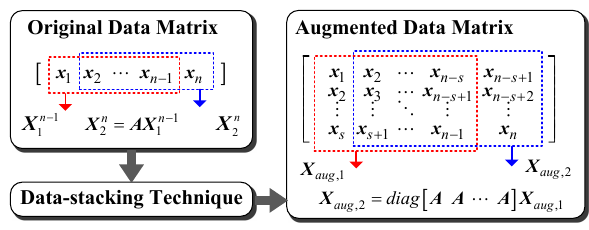}
\caption{Data-stacking technique — data matrices are augmented with higher row dimensions.}
\label{Fig4}
\end{figure}

 In DMD and DMDc algorithms, high dimensional data matrices are required to ensure an accurate solution containing all important modes. For the power electronic converter shown in Fig. 1, there are few measurable signals when the controller is unknown, resulting in insufficient rows in the data matrix. In order to increase the row dimension in the data matrix when there is a small number of measurement channels, a data-stacking technique based on the Hankel matrix \cite{r11} can be applied as shown in Fig. \ref{Fig4}.
 
After shift-stacking and time-delay operations, the augmented data matrices of the system states ${{\bf{X}}_{aug,1}},{\rm{  }}{{\bf{X}}_{aug,2}} \in {\mathbb{R}^{ms \times (n - s)}}$ are obtained with a higher row dimension than the original matrices ${\bf{X}}_1^{n - 1},{\bf{X}}_2^n \in {\mathbb{R}^{m \times (n - 1)}}$, where $s$ is the shift-stacking number. Similar processing can be performed on the input data matrix  ${\bf{U}}_1^{n - 1} \in {\mathbb{R}^{e \times (n - 1)}}$ to obtain ${{\bf{U}}_{aug}} \in {\mathbb{R}^{es \times (n - s)}}$. 

\section{Experimental Results}
\subsection{Experimental Setup}
\begin{figure}[htbp]
\centering
\includegraphics[width=0.95\columnwidth]{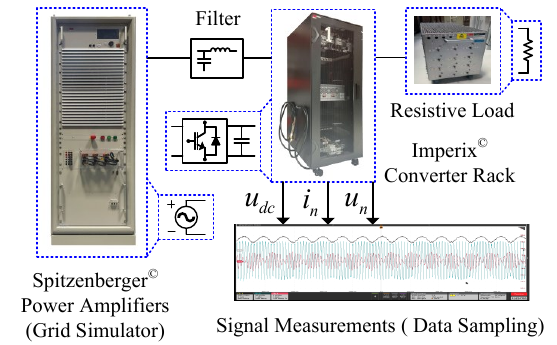}
\caption{Experimental setup and signal measurements of a single-phase converter in the electrified railway system.}
\label{Fig5}
\end{figure}

The DMD and DMDc algorithms are implemented based on the data collected from the equivalent experimental platform of a single-phase converter as shown in Fig. \ref{Fig5}, in which the topology is similar to that in Fig. 1, but system parameters are downscaled due to the available hardware resources, and main experimental parameters are summarized in Table \uppercase\expandafter{\romannumeral1}. The transient direct current control strategy \cite{r16} is used in the controller involving phase-locked loop (PLL), band-pass filter (BPF), DC-link voltage controller (DVC), and current controller (CC). However, it is worth mentioning that the measured signals are independent of the controller due to the assumption of the unknown controller structure in the proposed gray-box method.  

\begin{table}
\centering
\caption{Main Parameters of Experimental System}
\renewcommand{\arraystretch}{1.6}
\begin{tabular}{ccc} 
\hline\hline
\textbf{Parameter} & \textbf{Description}                                                                              & \textbf{Value}  \\ 
\hline
${u_g}$             &  Grid phase voltage (RMS)                                                               & 110  V        \\ \hline
${L_g}$             & Grid-side inductance                                                                      & 8  mH      \\ \hline
${L_n}$             & Converter-side inductance                                                                  & 4  mH        \\ \hline
${C_n}$             & Filter capacitance                                                                 & 10   uF        \\ \hline
${u_{dc}}$             & DC link voltage of the converter                                                   & 170  V        \\ \hline
${R_d}$             & DC link Load resistance                                                                     & 460 $\Omega$       \\ \hline
${C_d}$              & DC link support capacitance                                                                 & 800  uF          \\ \hline
${f_s}$               & Switching frequency                                                                         & 10  kHz       \\ \hline
${K_{BPF}}$             & Proportional gain of BPF                                                                  & 0.5         \\ \hline
${K_{pPLL}}$             & Proportional gain of PLL                                                                  & 0.1         \\ \hline
${K_{iPLL}}$             & Integral gain of PLL                                                                        & 5          \\ \hline
${K_{pVC}}$             & Proportional gain of DVC                                                                  & 0.09        \\ \hline
${K_{iVC}}$             & Integral gain of DVC                                                                        & 13         \\ \hline
${K_{pCC}}$              & Proportional gain of CC                                                                   & 0.01         \\ \hline
\hline
\end{tabular}
\end{table}

When low-frequency oscillations occur under low power and weak grid conditions \cite{r13}, the data is collected in a sampling window of 2 seconds with a frequency of 2500 Hz, such that $n = 5000$. For the DMD algorithm, all measured signals $({i_n}, {u_n}, {u_{dc}})$ are regarded as state signals, i.e., $ m = 3$, but for the DMDc algorithm, there are two state signals $({i_n}, {u_{dc}})$ and one input signal $({u_n})$ according to (1), i.e., $ m = 2, e = 1$. Through the data-stacking technique with the shift-stacking number $s$ = 1000, the dimension of state data matrices ${\bf{X}}_1^{n - 1},{\bf{X}}_2^n $ in DMD transform from $(3 \times 4999)$ to $(3000 \times 4000)$. For DMDc, the dimension of state data matrices ${\bf{X}}_1^{n - 1},{\bf{X}}_2^n $ transform from $(2 \times 4999)$ to $(2000 \times 4000)$, and the dimension of input data matrices ${\bf{U}}_1^{n - 1}$ transforms from $(1 \times 4999)$ to $(1000 \times 4000)$, highlighting the fact that the row dimensions of the data matrices are augmented significantly.

\subsection{Results}
\begin{figure}[htbp]
\centering
\includegraphics[width=0.95\columnwidth]{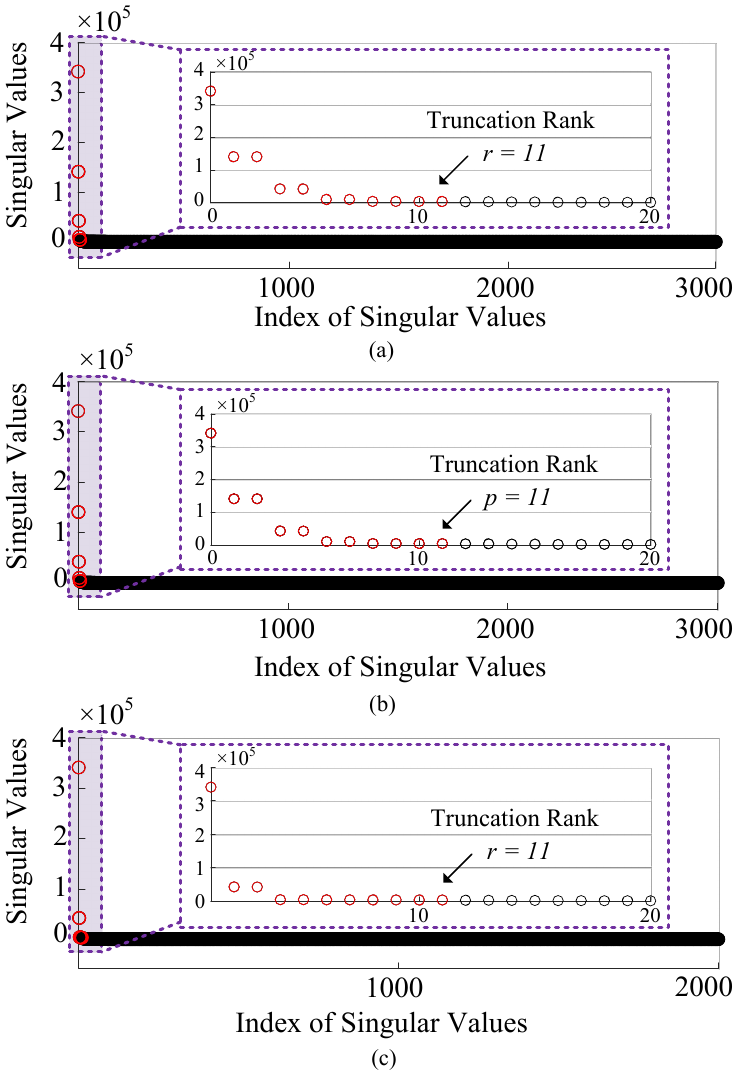}
\caption{Singular value decompositions (SVD) in DMD and DMDc algorithms and their zoom-in views: (a) SVD of data matrix ${\bf{X}}_1^{n - 1}$ in DMD, (b) SVD of composite data matrix ${\bf{\Omega}}_1^{n - 1}$ in DMDc, (c) SVD of data matrix ${\bf{X}}_2^n$ in DMDc — setting reduction ranks to 11 in all SVDs is reasonable.} 
\label{Fig6}
\end{figure}

Reducing the order of the dynamic matrix through SVD to improve computation efficiency is an important advantage of the DMD and DMDc algorithms, and it is necessary to determine the reduction rank of SVD to identify effective eigenvalues. It can be seen from Fig. \ref{Fig6} that the reduction ranks of the SVD in DMD and DMDc algorithms are set to 11, which are capable of holding most data information, and the remaining ineffective singular values can be disregarded because their values are approximately zero. On the other hand, it is reasonable to compare the two methods with the same reduction rank.

\begin{figure}[htbp]
\centering
\includegraphics[width=0.95\columnwidth]{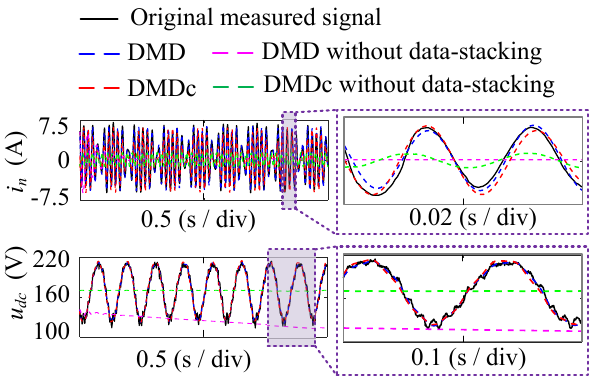}
\caption{Signals reconstruction with DMD and DMDc algorithms and their zoom-in views — data-stacking is necessary for algorithms implementation.} 
\label{Fig7}
\end{figure}

As shown in Fig. \ref{Fig7}, after carrying out necessary data-stacking, the reconstructed state signals performed by the black-box DMD and the proposed gray-box DMDc align closely with the original measured signals with the occurrence of oscillations at approximately 8.6 Hz, indicating that the correct implementation of both algorithms. However, measured signals cannot be reconstructed properly without data-stacking. 

\begin{figure}[htbp]
\centering
\includegraphics[width=0.9\columnwidth]{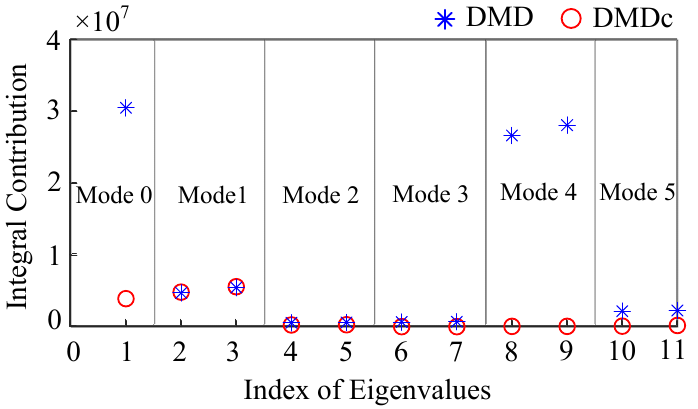}
\caption{Integral contribution of each eigenvalue extracted by DMD and DMDc algorithms — the contribution of each mode to the instability generation is presented.} 
\label{Fig8}
\end{figure}

\begin{figure}[htbp]
\centering
\includegraphics[width=0.95\columnwidth]{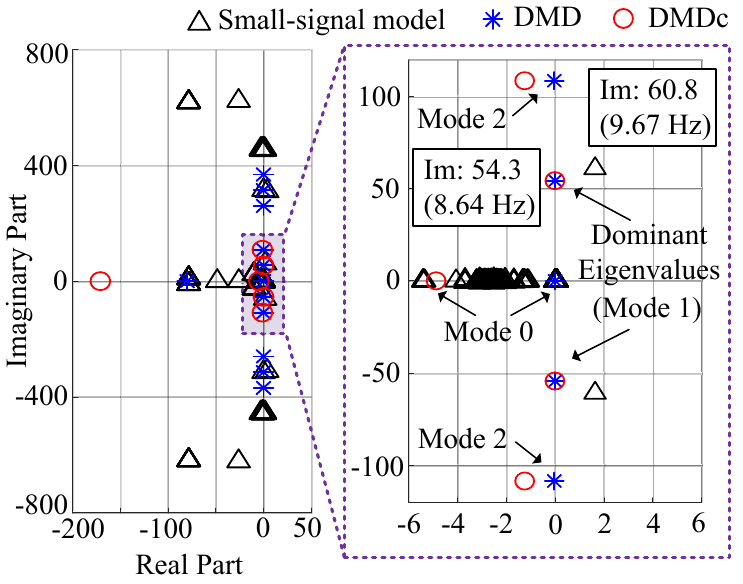}
\caption{Eigenvalues identification by DMD and DMDc algorithms and its zoom-in views — dominant eigenvalues identified by DMDc is more accurate than DMD.} 
\label{Fig9}
\end{figure}

Since the reduction rank of SVD is set to 11, there are 11 eigenvalues extracted by both black-box DMD and gray-box DMDc, and the integral contribution of each eigenvalue is calculated based on (13) as shown in Fig. \ref{Fig8}, representing the importance of each eigenvalue. Except for the basic eigenvalue (mode 0), other eigenvalues are presented in the form of conjugate pairs in the complex plane, and each pair of eigenvalues can be defined for an oscillation mode. Furthermore, the eigenvalues extracted by black-box DMD, gray-box DMDc, and small-signal mathematical models are compared in the complex plane as shown in Fig. \ref{Fig9}, all revealing that there is a pair of dominant eigenvalues with positive (or around zero) real parts, which corresponds to mode 1 with negative (or critical) damping, indicative of oscillations at around 9 Hz. It is worth mentioning that overmodulation limits the divergence of the oscillation amplitude, resulting in a real part of the dominant eigenvalues obtained from the data-driven methods being around zero. 

Hence, both black-box DMD and gray-box DMDc can identify dominant eigenvalues and reveal the significant oscillation mode with underlying damping and frequency for stability analysis, but DMDc gives more accurate information. To be specific, DMDc only considers mode 1 to be dominant, but DMD misjudges mode 4 to be more important, as shown in Fig. \ref{Fig8}. Furthermore, Fig. \ref{Fig9} depicts that the dominant eigenvalues identified by DMDc match the small-signal model and oscillation waveforms better, while DMD gives another pair of additional interference eigenvalues (mode 2) also showing critical negative damping, but it is not recognized by the small-signal model, and the oscillations with corresponding frequency do not exist in the measured signal waveform. 

\begin{figure}[htbp]
\centering
\includegraphics[width=0.9\columnwidth]{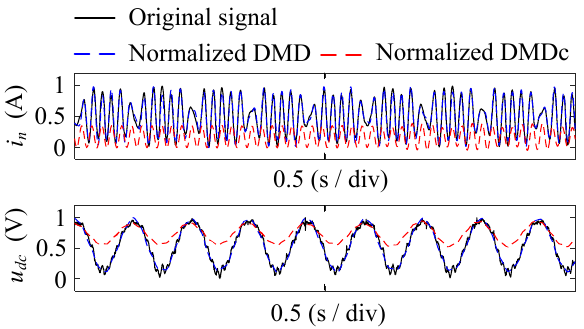}
\caption{Comparison of black-box DMD and gray-box DMDc after data normalization — gray-box DMDc is more sensitive to data structure but has stronger interpretability.} 
\label{Fig10}
\end{figure}

In addition, the proposed gray-box DMDc is more sensitive to the physical relationship between the data provided by the state equations in (1), which can be verified as shown in Fig. \ref{Fig10}. If the data is normalized in advance, the physical relationship between the variables in (1) will no longer hold, leading to a significant error in the gray-box DMDc algorithm, while black-box DMD is not affected. Rather than a drawback, this property highlights the enhanced interpretability of DMDc and renders the analysis results more reliable with the inherent physical significance.

\section{Conclusions}

A gray-box stability analysis mechanism based on dynamic mode decomposition with control (DMDc) algorithm is proposed for the power electronic converters, and the data-stacking technique is used for the dimension augmentation of data matrices to meet the algorithmic requirements. The system state matrix and input matrix are approximated to identify dynamic modes and eigenvalues by combining the DMDc algorithm and the physical constraint from converter state equations. Compared with the conventional DMD method, the identified dominant eigenvalues can match the small-signal model and measured oscillation waveforms more accurately. Moreover, the reconstruction of measured signals is more sensitive to the physical meaning of the data, showing the analysis results are more reliable with interpretability. The proposed gray-box DMDc method can provide more accurate and interpretable stability monitoring and oscillation identification for converters with partially known circuit structures but unknown controllers. To extend the scope of this article in the future, there can be more centered discussions in terms of the physical constraints of the system, such as methods to tackle the modeling inaccuracy of state equations considering the nonlinearity.

\bibliographystyle{IEEEtran}
\bibliography{Mybib}

\end{document}